\documentclass{article}
\usepackage{spconf,amsmath,epsfig}

\title{Disentangled Non-Local Network for Hyperspectral and \\ 
LiDAR Data Classification}

\name{Wenxia Liu$^{1,2}$, Feng Gao$^{1,2}$, Junyu Dong$^{1,2,*}$}
\address{$^1$College of Information Science and Engineering, Ocean University of China \\
$^2$ Institute of Marine Development, Ocean University of China
\thanks{This work was supported in part by the National Key Research and Development Program of China under Grant 2018AAA0100602, in part by the
National Natural Science Foundation of China under Grant U1706218, and in part by the Key Research and Development Program of Shandong Province
under Grant 2019GHY112048. (Email: dongjunyu@ouc.edu.cn)}}

\begin{document}

\maketitle

\begin{abstract}
As the ground objects become increasingly complex, the classification results obtained by single source remote sensing data can hardly meet the application requirements. In order to tackle this limitation, we propose a simple yet effective attention fusion model based on Disentangled Non-local (DNL) network for hyperspectral and LiDAR data joint classification task. In this model, according to the spectral and spatial characteristics of HSI and LiDAR, a multiscale module and a convolutional neural network (CNN) are used to capture the spectral and spatial characteristics respectively. In addition, the extracted HSI and LiDAR features are fused through some operations to obtain the feature information more in line with the real situation. Finally, the above three data are fed into different branches of the DNL module, respectively. Extensive experiments on Houston dataset show that the proposed network is superior and more effective compared to several of the most advanced baselines in HSI and LiDAR joint classification missions.
\end{abstract}


\section{Introduction}
\label{sec:intro}

With the advancement of spectral imaging technology, hyperspectral images (HSIs) have received increasing attention, and have been employed in many applications, such as marine organism monitoring \cite{b1} and land cover classification \cite{b2}. On the other hand, LiDAR data is also widely used in the classification of high-value crops \cite{b3}, urban land use analysis. Hyperspectral and LiDAR data grow rapidly in recent years, inducing many researchers to work on the joint land cover classification from both data.

HSI has the characteristics of high spectral space dimension and strong data correlation. Meanwhile, LiDAR images have high spatial resolution, but the spectral information of it is not as rich as that of HSI. Therefore, when performing classification tasks, if one single mode of data is used, the classification performance may subject to certain limitations. For example, due to the lack of elevation information or the influence of various spectral changes, a single HSI can hardly distinguish the difference between grasses in different areas \cite{b4}. Therefore, using multimodal data for joint classification allows people to more accurately identify the interested ground objects.

Until now, many works have successfully demonstrated the advantages of using multimodal data for classification. For example, the fusion of HSI and multispectral imagery can greatly improve the spatial resolution \cite{b5}. Chen {\it et al.} \cite{b6} proposed a new feature fusion framework based on deep neural network (DNN), which can effectively extract features from HSI and LiDAR data, and both features are fused through fully connected DNNs.

In particular, with the recent advancement of convolutional neural networks (CNNs), the fusion strategy based on CNN begins to prevail. In  \cite{b7}, an end-to-end layered fusion module for CNN is proposed, enabling HSI and LiDAR data to achieve good collaborative classification performance. Although the module shows its excellent performance on the corresponding datasets, the larger receiving field in the convolutional layer and the existence of pooling layer lead to the lower spatial resolution in the deeper CNN layer, and the multiscale information of HSI are not fully applied \cite{b8}.  Therefore, it is difficult to describe the significant boundary of fine objects in classification, and correspondingly, the prediction category tends to be spatially dispersed.

To solve the above problem, inspired by Yin's work \cite{b9}, we developed a simple yet effective HSI and LiDAR joint classification method based on distangled non-local (DNL) network. The traditional dot-product based attention is split into a whitened pairwise term and a unary term. The pairwise term fuses the hyperspectral and LiDAR features. The unary term captures the influence of one pixel generally over all the pixels in hyperspectral data. The experimental results on the Houston dataset demonstrate the effectiveness of the proposed method.

\begin{figure}[h]
\begin{center}
\includegraphics [width=2in]{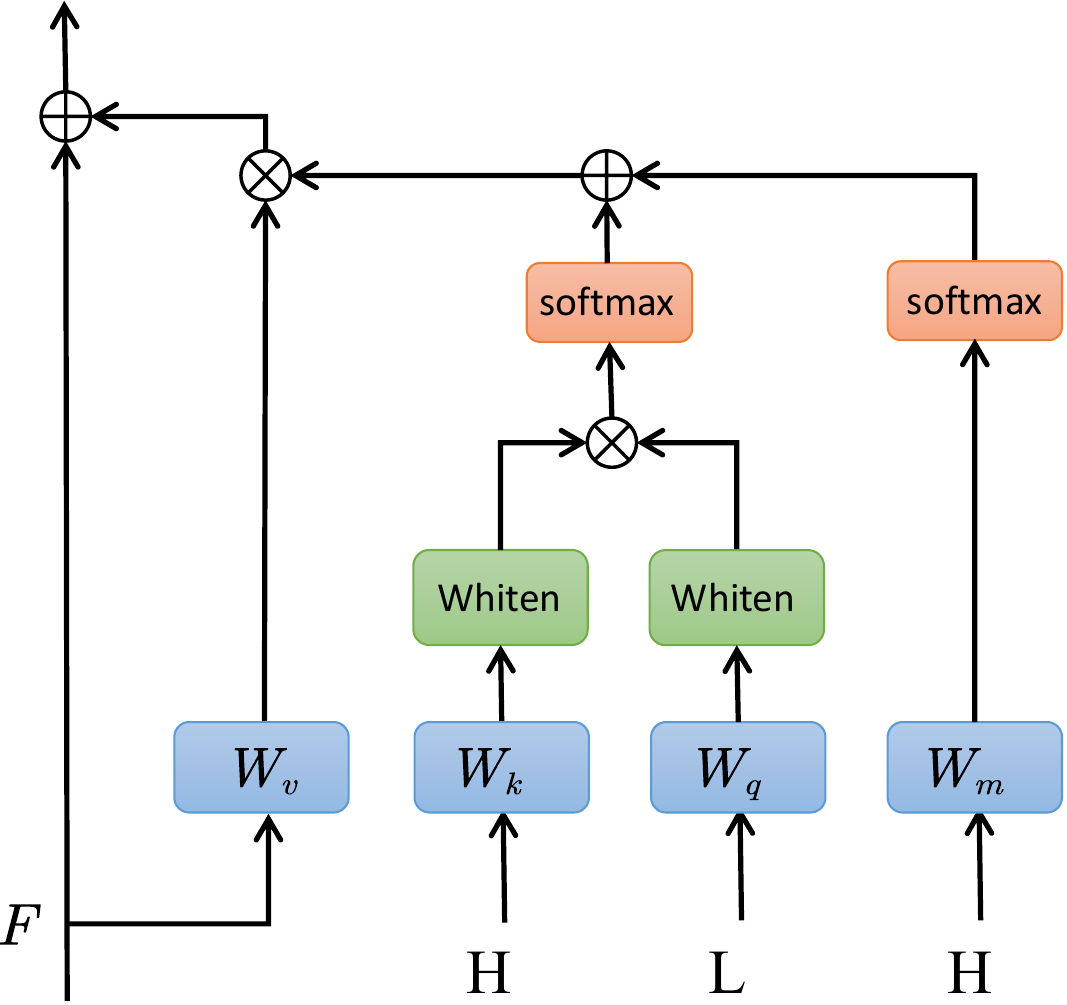}
\caption{Schematic illustration of the Disentangled Non-Local (DNL) module. $F$ represents the mixed feature obtained by adding HSI feature and LiDAR feature according to a certain strategy, while H and L represent the HSI features and LiDAR features, respectively.}
\end{center}
\end{figure}

\section{Methodology}

\subsection{Feature Extraction}

Extraction of features from HSI using residual network. To extract multiscale features, $1\times1$, $3\times3$, $5\times5$ and $7\times7$ filters are mixed for depth convolution operation. Meanwhile, the number of these four types of filters is equal, which is one fourth of the number of channels for input features. Specifically, one quarter of the channels are convolved with the $1\times1$ filter, a quarter of the channels are convolved with the $3\times3$ filter, and so on. As such, space correction can be learned at different scales. Each residual block can be expressed as
\begin{equation}
P(X) = s( f(X) + g(X) ),
\end{equation}
where $X$ denotes the input of the residual block, and $P(X)$ denotes the output of the residual block. $s(\cdot)$ represents the ReLU activation function. In addition, $f(X)$ represents the residual learning function, which performs a series of nonlinear transformation operations, and obtains the output of batching operations. $g(X)$ is a single mapping function, which can alleviate the problem of gradient disappearance in the network.

At the same time, LiDAR data extracted its elevation information and other features through CNN. After that, the processed HSI and LiDAR features were converted to the same scale for addition operation, and then several pooling and convolution operations were performed to obtain the mixed feature $F$.

\subsection{Joint classification based on DNL}

Non-local blocks can enhance the context modeling capability of conventional CNNs. However, according to Yin's work \cite{b10}, the attention computation of non-local can be divided into two terms, namely, paired terms and unary terms, which are tightly coupled and thus affect each other's learning. Disentangled non-local (DNL) attention can eliminate this dilemma. As shown in Fig. 1, HSI and LiDAR features together with $F$ are used as the input of DNL to fuse multimodal information. On the basis of not affecting the contextual relationship, the pixel-wise relationship can be learnt within the same category of regions, and the influence from boundary pixels to all image pixels can also be learnt effectively.

The DNL module counts pairs between the features of two locations to catch long-distance dependencies. Let $X_m$ denotes the input features at position $m$, the output features $Y_m$ of the DNL module can be computed as
\begin{equation}
Y_m = \sum_{n \in \Omega}M(X_m,X_n)g(X_n),
\end{equation}
where $M$ represents the set of all pixels on the feature map of $H\times W$, and $g(\cdot)$ is the value transformation function with parameter $W_v$. ($X_m$, $X_n$) is the embedding similarity function from pixel $n$ (called key pixel) to pixel $m$ (called query pixel), which can be expressed as
\begin{equation}
M(X_m , Y_n) =\sigma(q_m^T k_n),
\end{equation}
\begin{equation}
\sigma(q_m^T k_n) = \sigma(\underbrace{u_q^T k_n}_{unary} + \underbrace{(q_m - u_q)^T (k_n - u_k)}_{pairwise}),
\end{equation}
where $q_m$ represents the query embedding of pixel $m$, and $k_n$ represents the key embedding of pixel $n$. $\sigma(\cdot)$ represents the softmax function. The first term represents the unary relation in which the key pixel $n$ has the same influence on all query pixel $m$. The second term represents  the pure pairwise relation between the query pixel $m$ and key pixel $n$.

\begin{table}[htbp]
\caption{Classification results corresponding to different inputs. L, H and $F$ represent LiDAR features, HSI features and fused features, respectively.}
\begin{center}
\begin{tabular}{|c|c c c c|c|}
\hline
NO. &  $W_v$ & $W_k$ & $W_q$ & $W_m$ & OA \\
\hline \hline
1 &$F$ & H & H & H & 93.48$\pm$0.21 \\
2 &$F$ & H & L & L & 93.14$\pm$0.45\\
3 &$F$ & L & L & L & 92.85$\pm$0.52 \\
4 &H & H & L & H & 93.12$\pm$0.63 \\
5 &L & H & L & H & 92.71$\pm$0.46 \\
6 &$F$& H & L & H & 87.10$\pm$0.64 \\
7 &$F$ & H & L & H & 69.11$\pm$0.72 \\
8 &L & L & L & L & 46.96$\pm$0.25 \\
9 &H & H & H & H & 88.52$\pm$0.34 \\
10 &$F$ & H & L & H & 93.74$\pm$0.76 \\
\hline
\end{tabular}
\label{tab1}
\end{center}
\end{table}

\begin{table*}[htbp]
\small
    \centering
\caption{Comparison of the classification accuracy (\%) using the Houston data}
\begin{center}
\begin{tabular}{|c c|c c c c c c|}
\hline
NO. & Class & SVM & CNN-PPF & CRNN & CNN-MRF & HRWN & DNL \\
\hline\hline
1 & Health grass $\left(198/1053\right)$ & 82.43$\pm$6.95 & 83.57$\pm$0.42 & 83.00$\pm$0.23& 85.77$\pm$0.31  & 85.61$\pm$0.09 &82.71$\pm$0.22  \\
2 & Stressed grass $\left(190/1064\right)$ & 82.05$\pm$1.31 & 98.21$\pm$0.15 & 79.41$\pm$0.98 & 86.28$\pm$0.45 & 85.17$\pm$1.25 &99.62$\pm$1.12\\
3 & Synthetic grass $\left(192/505\right)$ &99.80$\pm$0.04 & 98.42$\pm$0.44 & 99.8$\pm$0.12 & 99.00$\pm$0.02 & 99.57$\pm$0.18 &99.60$\pm$0.12 \\
4 &Tress $\left(188/1056\right)$ &92.80$\pm$3.84 & 97.73$\pm$0.27 & 90.15$\pm$0.78 &92.85$\pm$0.87 & 92.20$\pm$0.62 & 95.45$\pm$0.23\\
5 &Soil $\left(186/1056\right)$ &98.48$\pm$0.97 & 96.50$\pm$1.03 &99.71$\pm$0.12 &100.0$\pm$0.00 &100.0$\pm$0.00 &99.90$\pm$0.43\\
6 &Water $\left(182/143\right)$ &95.10$\pm$2.06 & 97.20$\pm$2.29 &83.21$\pm$5.44 & 98. 15$\pm$0.25 & 98.15$\pm$0.86 &94.40$\pm$0.53\\
7 &Residential $\left(196/1072\right)$ &75.47$\pm$0.75 &85.82$\pm$0.59 &88.06$\pm$1.76 & 91.64$\pm$0.38  &95.98$\pm$0.89 &88.71$\pm$1.23\\
8 &Commercial $\left(191/1036\right)$ & 46.91$\pm$8.43 & 56.51$\pm$0.58& 88.61$\pm$2.37 & 80.79$\pm$0.56 &97.59$\pm$0.52 &94.77$\pm$1.45\\
9 &Road $\left(193/1059\right)$ &77. 53$\pm$2.76 & 71.20$\pm$1.04 & 66.01$\pm$9.32 & 91.37$\pm$2.08 & 88.66$\pm$0.63 & 97.35$\pm$0.54\\
10 &Highway $\left(191/1036\right)$ &60.04$\pm$1.05 & 57.12$\pm$1.82 & 52.22$\pm$7.84 & 73.35$\pm$0.84 & 86.23$\pm$0.34 &81.99$\pm$0.42\\
11 & Railway $\left(181/1054\right)$ &81.02$\pm$2.26 & 80.55$\pm$0.36 & 81.97$\pm$3.02 & 98.87$\pm$1.33 & 97.98$\pm$3.06 & 96.77$\pm$0.26\\
12 &Parking lot 1 $\left(192/1041\right)$ &85.49$\pm$2.99 &62.82$\pm$2.64 & 69.83$\pm$7.64 &89.38$\pm$2.71 &97.40$\pm$0.44 &99.71$\pm$0.53\\
13 &Parking lot 2 $\left(184/285\right)$ &75.09$\pm$0.97 &63.86$\pm$0.77 & 79.64$\pm$5.13 &92.75$\pm$1.30 &91.47$\pm$1.25& 90.52$\pm$1.84\\
14 &Tennis court $\left(181/247\right)$ & 100.0$\pm$0.00 & 100.0$\pm$0.00 &100.0$\pm$0.00 & 100.0$\pm$0.00 &100.0$\pm$0.00 &100.0$\pm$0.00 \\
15 &Running track $\left(187/473\right)$ &98.31$\pm$0.14 & 98.10$\pm$0.39 & 100.0$\pm$0.00 &100.0$\pm$0.00 &100.0$\pm$0.00 &100.0$\pm$0.00\\ \hline\hline
 & OA &80.49$\pm$0.38 & 83.33$\pm$0.70 &88.55$\pm$0.71  &90.61$\pm$0.54 &93.61$\pm$0.75 & 93.74$\pm$0.76\\
 & AA &83.37$\pm$1.10 &83.21$\pm$0.33 &90.30$\pm$0.64 &92.01$\pm$0.48 &94.40$\pm$0.63 &94.78$\pm$0.62\\
 & Kappa &78.98$\pm$0.45 &81.88$\pm$0.77 &87.56$\pm$0.77 & 89.87$\pm$0.58 &93.09$\pm$0.81 &93.41$\pm$0.79\\
\hline
\end{tabular}
\label{tab2}
\end{center}
\end{table*}

\section{Experimental Results and Analysis}
\subsection{Experimental Setup}

To verify the effectiveness of the proposed disentangled non-local network for HSI and LiDAR joint classification, we used Houston dataset acquired by the NSF-funded Center in June 2012 over the University of Houston campus and neighboring areas. There are 144 spectral bands that range from 0.38 to 1.05 $\mu$m in it. Beyond that, it presents an area with the size of $349\times1905$ pixels and the spatial resolution of 2.5 m. The specific number of training and testing samples can is illustrated in Table \ref{tab2}.

\subsection{Results and Analysis}

We test different feature combinations for the proposed DNL. As illustrated in Table \ref{tab1}, classification using LiDAR image alone is unsatisfactory. Similarly, classification using HSI alone is less effective. The best results are achieved when we use the HSI feature as the key, the LiDAR feature as the query, and the fused feature as the value. Thereby, we use such configurations in our following experiments.

To verify the validity of the proposed network, the classification performance of the proposed DNL was compared with five closely related methods, including SVM, CNN-PPF\cite{b11}, CRNN\cite{b12}, CNN-MRF\cite{b13}, HRWN\cite{b14}. Three evaluation metrics are used, including overall accuracy (OA), average accuracy (AA) and kappa coefficient (Kappa), as shown in Table \ref{tab2}. It can be seen that the proposed DNL shows superiority over other methods, which implies that the disentangled non-local mechanism indeed performs well in exploiting the multimodal feature correlations.

\begin{table}[htbp]
\caption{The comparison between Non-Local (NL) and DNL networks.}
\begin{center}
\begin{tabular}{|c|ccc|}
\hline
Method & OA &  AA & Kappa\\
\hline \hline
NL & 93.22$\pm$0.68 &93.99$\pm$0.64 &93.01$\pm$0.72 \\
DNL& 93.74$\pm$0.76 &94.78$\pm$0.62 &93.41$\pm$0.79 \\
\hline
\end{tabular}
\label{tab3}
\end{center}
\end{table}

Beyond that, we also present the results of using traditional non-local (NL) and DNL in HSI and LiDAR classification, as shown in Table 3. The results show that DNL performs better than that of traditional NL, which exactly confirms the analysis in Section 2. Fig. 2 illustrates the classification results of the proposed method, from which we can clearly see that the boundary of the network using DNL is clearer, and more clear regional cues can be learnt.

\begin{figure*}[ht]
\centering
\includegraphics [width=6.5in]{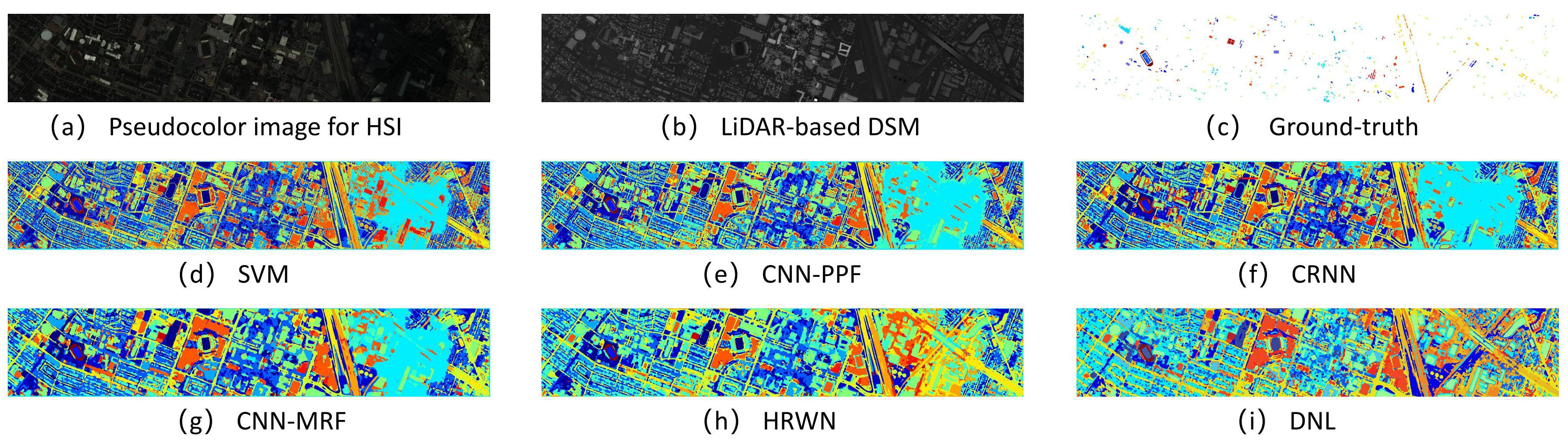}
\caption{Classification maps for the Houston data obtained with different methods.}
\end{figure*}

\section{Conclusion}

In this paper, we propose a novel DNL network for HSI and LiDAR joint classification. In the network, we first extract the corresponding features from HSI and LiDAR images according to their own characteristics, respectively. Afterwards, in order to obtain the feature information more appropriate to the actual environment, the extracted HSI and LiDAR feature information is obtained through some operations to obtain the anisotropic information, so as to prepare for the fusion of the latter two types of feature information. Finally, the above three types of feature information are fed into the DNL model to obtain more accurate contextual dependence. Experiments show that the performance of the proposed model is superior to several closely related methods.

\end{document}